\documentclass[aps,pra,twocolumn,superscriptaddress,floatfix,showpacs]{revtex4-2}

\usepackage{graphicx}
\usepackage{amsmath,amssymb,amsfonts}
\usepackage{bm}
\usepackage{dsfont}
\DeclareMathOperator{\sech}{sech}
\usepackage{color}
\usepackage[colorlinks=true,
            citecolor=blue,
            linkcolor=blue,
            urlcolor=blue]{hyperref}

\newcommand{\rev}[1]{#1}
\newcommand{\ket}[1]{\left|#1\right\rangle}
\newcommand{\bra}[1]{\left\langle#1\right|}

\begin{document}

\title{Investigating entanglement dynamics with dynamical decoupling together with initial qubit-reservoir correlations}

\author{Rifzah Shahid}
\affiliation{Department of Physics, Quaid-i-Azam University, Islamabad 45320, Pakistan}

\author{Amna Qadeer}
\affiliation{Department of Physics, Quaid-i-Azam University, Islamabad 45320, Pakistan}

\author{Ali Raza Mirza}
\email{ali.mirza@qau.edu.pk}
\affiliation{Department of Physics, Quaid-i-Azam University, Islamabad 45320, Pakistan}
\affiliation{National Centre for Physics, Islamabad 45320, Pakistan}


\begin{abstract}
We study the exact entanglement dynamics of two qubits coupled to a common bosonic reservoir through a pure dephasing interaction. We ask how this dynamics changes when the qubits and the reservoir already share correlations before the qubit state is prepared. In that case the system and reservoir first interact for a long time and settle into a joint equilibrium state. Only afterward is the qubit pair projected onto the desired state by a local measurement. For an Ohmic spectral density we obtain the exact reduced two qubit dynamics. Using the Wootters concurrence we show that initial correlations always speed up entanglement sudden death relative to the factorized case, and that this gap grows quadratically with the coupling strength in the weak-to-moderate coupling regime. Finally we show that a symmetric sequence of periodic $\pi$ pulses applied to both qubits extends the entanglement lifetime close to a factor of seven over the range of pulse numbers studied.
\end{abstract}

\maketitle

\section{Introduction}
\label{sec:intro}

Every real quantum system sits inside an environment. Understanding how that environment destroys quantum coherence and entanglement is central to building quantum technologies~\cite{BreuerPetruccione2007,Weiss2008}. Most treatments of open systems make two simplifying assumptions. First they assume the coupling between system and environment is weak enough for a perturbative master equation such as Lindblad to apply. Second they assume the initial system and environment state is a simple product state in which the environment sits in its own equilibrium, uncorrelated with the system. This second assumption of no initial correlations is convenient. But it is increasingly seen as unjustified once the coupling is not negligible. A system that has interacted with its reservoir for any real length of time will generally become correlated with it before any later state preparation or measurement takes place~\cite{RomeroPaz1997,Lutz2003,BanerjeeGhosh2003,SmirneBreuerPiiloVacchini2010,DijkstraTanimura2010,MorozovMatheySRopke2012,ChaudhryGong2013a,ChaudhryGong2013b,ChaudhryGong2013c,BuserCerrilloSchallerCao2017,deVegaAlonso2017}.

There is a natural and physically transparent way to build in these correlations. Imagine that the full Hamiltonian of system plus environment, not the environment Hamiltonian alone, governs the state before the system is prepared. The joint system and environment are allowed to reach the equilibrium state $e^{-\beta H}/Z$. Only afterward is the desired system state obtained through a projective measurement or unitary operation. Because the projection acts after system and environment have already equilibrated together, the resulting reduced environment state keeps an imprint of the interaction. The later open system dynamics can then differ, sometimes drastically, from what a naive factorized initial state would predict \cite{mirza2021master}. This projected thermal construction has already been used to solve a single dephasing two level system essentially exactly~\cite{ChaudhryGong2013a,ChaudhryGong2013b,MorozovMatheySRopke2012}. It has been written as a master equation that directly incorporates the correlations in the joint equilibrium state~\cite{mirza2021master}. It has also been extended to spin environments and to the estimation of spin reservoir parameters~\cite{MajeedChaudhry2019,mirza2024role,mirza2024impact}. And it has been shown to substantially sharpen the precision with which environmental parameters can be estimated from single and two qubit quantum probes, including through periodic control pulse sequences and alternative unitary state preparation protocols~\cite{AtherChaudhry2021,mirza2024improving,mirza2023improving,mirza2025effect,abbas2025sensing}.

Beyond single-qubit observables, extensive effort has been devoted to understanding the open-system dynamics of multi-partite quantum correlations, where environmental coupling often leads to sudden death or non-monotonic revival of entanglement~\cite{YuEberly2004,Bellomo2007}. In this regime, dynamical decoupling (DD) traditionally employed to suppress single-qubit dephasing and relaxation~\cite{ViolaLloyd1998,Uhrig2007} has emerged as a powerful control protocol for protecting and steering bipartite entangled states against local and correlated noise baths~\cite{Mukherjee2013,Addis2015}. Rather than merely freezing decay, pulse sequences applied to entangled nodes can engineer effective filter functions that selectively decouple destructive environmental spectral modes while leveraging non-Markovian memory effects to prolong or restore nonlocal correlations. But entanglement, a genuinely two body resource, is often the quantity that matters most for quantum information applications. It is not obvious in advance whether system-environment initial correlations help or hurt the survival of entanglement between two qubits sharing a common bath. Reference~\cite{MajeedChaudhry2019} addressed this question for two qubits coupled to a spin environment. It found that initial correlations can substantially change entanglement sudden death and sudden birth. The size and even the sign of the effect depend sensitively on microscopic details of the environment: the relative strength of the qubit environment coupling, the environment's internal energy scales, and any interactions among the environment spins. It is natural to ask whether a similar, and more tractable, story holds for the complementary and widely used case of a common bosonic dephasing environment, where the single qubit projected thermal construction is already exactly solvable~\cite{ChaudhryGong2013a}. It is equally natural to ask whether the dynamical decoupling strategy is effective at protecting two qubit entanglement.

In this paper we answer both questions. We consider two qubits coupled identically to a common Ohmic bosonic reservoir through a pure dephasing interaction, the natural two qubit generalization of the independent boson model. We solve exactly for the reduced two qubit density matrix under both the factorized and the projected thermal (correlated) initial conditions. Because the interaction commutes with the qubit Hamiltonian, the model stays exactly solvable to all orders in the coupling for both preparations. This lets us isolate the effect of the initial correlations from any uncontrolled approximation. We use the Wootters concurrence to quantify the entanglement dynamics. We obtain a closed form expressions for entanglement sudden death (ESD) for both the factorized case and the correlated case. We show two main results. First, initial correlations always hasten entanglement sudden death, and the gap between the correlated and factorized cases grows faster than linearly with the coupling strength. Second, a symmetric sequence of periodic $\pi$ pulses substantially prolongs the entanglement lifetime for both preparations. This is a two qubit realization of the dynamical decoupling strategies used for parameter estimation in Ref.~\cite{AtherChaudhry2021} and developed more broadly in the qubit noise spectroscopy and dynamical decoupling literature~\cite{ViolaLloyd1998,ViolaKnillLloyd1999,NorrisPazSilvaViola2016,deLangeWangRisteDobrovitskiHanson2010,PokharelAnandFortmanLidar2018}. It acts as a nearly independent knob on top of the correlation induced physics.

The rest of the paper is organized as follows. Section~\ref{sec:model} presents the Hamiltonian and states the exact reduced two qubit dynamics. It covers the factorized initial state first (Sec.~\ref{sec:uncorrelated}) and then the correlated, projected thermal initial state (Sec.~\ref{sec:correlated}). Full derivations for both cases are left to Appendices~\ref{app:uncorrelated} and~\ref{app:correlated} so the main text can stay focused on physical content. Section~\ref{sec:entanglement} uses these results to study entanglement sudden death with and without initial correlations. Section~\ref{sec:control} shows how periodic $\pi$ pulse control fields protect the entanglement of both preparations. We conclude in Sec.~\ref{sec:conclusion}.

\section{Model: two qubits in a common bosonic reservoir}
\label{sec:model}

We consider two qubits coupled identically (``symmetrically'') to a common bath of harmonic oscillators through their $\sigma_z$ operators. Setting $\hbar=k_B=1$ throughout, the total Hamiltonian is $H=H_S+H_B+H_I$, with
\begin{align}
H_S &= \frac{\omega_0}{2}\left(\sigma_z^{(1)}+\sigma_z^{(2)}\right), \\
H_B &= \sum_r \omega_r\, b_r^\dagger b_r, \\
H_I &= J_z \sum_r \left(g_r^* b_r + g_r b_r^\dagger\right), \qquad J_z \equiv \sigma_z^{(1)}+\sigma_z^{(2)},
\label{eq:HI}
\end{align}
where $\omega_0$ is the common qubit splitting, $b_r,b_r^\dagger$ are bosonic ladder operators for reservoir mode $r$ of frequency $\omega_r$, and $g_r$ is its coupling strength to the collective qubit operator $J_z$. $H_S$ and $H_I$ are both built from $J_z$, so they commute: $[H_S,H_I]=0$. This is a pure dephasing coupling. The qubit populations in the joint $\sigma_z^{(1)}\otimes\sigma_z^{(2)}$ eigenbasis are exact constants of motion, and no energy passes between the qubits and the reservoir. All the reservoir does is randomize the relative phases between qubit basis states. This commutation is also what makes the model exactly solvable. It forces every Magnus expansion term of order three and higher to vanish, so the formally infinite Dyson series for the propagator truncates exactly at second order, for any coupling strength (Appendix~\ref{app:uncorrelated}).

Because both qubits couple to the same reservoir modes, the reservoir cannot tell which qubit imprinted a given phase. It therefore mediates an effective correlation between the two qubits. This is the key structural feature that lets a common environment generate entanglement, not just destroy it, unlike two independent local reservoirs. Throughout we take the reservoir spectral density to be Ohmic with an exponential cutoff,
\begin{equation}
J(\omega) = g\,\omega\, e^{-\omega/\omega_c},
\label{eq:Jomega}
\end{equation}
with $g$ a dimensionless coupling strength and $\omega_c$ the high frequency cutoff.

\subsection{Dynamics without initial correlations}
\label{sec:uncorrelated}

The standard, and simplest, choice of initial condition is a factorized system-reservoir state,
\begin{equation}
\rho(0) = \rho_S(0)\otimes\rho_B, \qquad \rho_B = \frac{e^{-\beta H_B}}{Z_B},
\label{eq:factinit}
\end{equation}
with the reservoir in its own thermal equilibrium state at inverse temperature $\beta=1/T$ and $\rho_S(0)$ the arbitrary qubit state we want to prepare. Physically this amounts to switching the qubits on to the reservoir only at $t=0$, with no shared history beforehand. That idealization is exact only in the limit of vanishing system-environment interaction. Carrying out the Magnus expansion calculation detailed in Appendix~\ref{app:uncorrelated}, we find the exact reduced density matrix is
\begin{widetext}
\begin{equation}
[\rho_S(t)]_{k'l';kl} = [\rho_S(0)]_{k'l';kl}\;
e^{-i\frac{\omega_0}{2}(k'+l'-k-l)t}\;
e^{-i\frac{\Phi(t)}{2}(k'l'-kl)}\;
e^{-\frac14(k+l-k'-l')^2\Gamma(t)},
\label{eq:rhoSfinal}
\end{equation}
\end{widetext}
in the joint $\sigma_z^{(1)},\sigma_z^{(2)}$ eigenbasis $\ket{k,l}$, $k,l=\pm1$, which we identify with the computational basis via $+1\leftrightarrow\ket0$, $-1\leftrightarrow\ket1$. Two real functions of time control everything. One is a collective dynamical phase $\Phi(t)$, which multiplies the eigenvalue difference $k'l'-kl$ of the bath-mediated operator $\sigma_z^{(1)}\sigma_z^{(2)}$ and does not decay; it only shifts the phase of coherences that connect states of different $\sigma_z^{(1)}\sigma_z^{(2)}$ parity. The other is the decoherence function
\begin{equation}
\Gamma(t) = 4\int_0^{\infty} d\omega\; \frac{J(\omega)}{\omega^2}\,\big[1-\cos(\omega t)\big]\,\coth\!\left(\frac{\beta\omega}{2}\right),
\label{eq:Gamma}
\end{equation}
which is the familiar finite temperature spin boson dephasing integral, dressed here by the collective factor $\tfrac14(k+l-k'-l')^2\in\{0,1,4\}$. Coherences connecting states whose $J_z$ eigenvalues differ by 4 (for example $\vert{}00\rangle \leftrightarrow \vert{}11\rangle$) decay four times faster in the exponent than coherences whose $J_z$ eigenvalues differ by 2 (for example $\vert{}00\rangle \leftrightarrow \vert{}01\rangle$). The coherence between $\vert{}01\rangle$ and $\vert{}10\rangle$ shares the same $J_z$ eigenvalue and the same $\sigma_z^{(1)}\sigma_z^{(2)}$ eigenvalue, so it never decays and never rotates at all: $\{\vert{}01\rangle,\vert{}10\rangle\}$ forms a decoherence free subspace, protected because the common reservoir cannot tell the two qubits apart. Rather than the separable product state used in earlier work on this model, we take the initial qubit state to be the maximally entangled Bell state
\begin{equation}
\ket{\psi}=\ket{\Phi^+}=\frac{1}{\sqrt2}\bigl(\ket{00}+\ket{11}\bigr),
\label{eq:bellstate}
\end{equation}
so that $\rho_S(0)=\ket{\Phi^+}\bra{\Phi^+}$ has nonzero entries only at the four index combinations $(k',l'),(k,l)\in\{(+1,+1),(-1,-1)\}$,
\begin{equation}
\rho_S(0)=\frac12
\begin{pmatrix}
1&0&0&1\\
0&0&0&0\\
0&0&0&0\\
1&0&0&1
\end{pmatrix}
\qquad(\text{basis }\ket{00},\ket{01},\ket{10},\ket{11}).
\label{eq:rho0}
\end{equation}
Because Eq.~\eqref{eq:rhoSfinal} only multiplies each initial matrix element by a $c$-number, it can never create population where none existed initially, so $\rho_S(t)$ keeps exactly the same sparsity pattern at every later time: the populations of $\ket{00}$ and $\ket{11}$ stay frozen at $\rho_{11}(t)=\rho_{44}(t)=\tfrac12$, the $\{\ket{01},\ket{10}\}$ block stays identically zero, and only the single coherence $\rho_{14}(t)=\rho_{41}(t)^*$ evolves. For the diagonal elements every exponent in Eq.~\eqref{eq:rhoSfinal} vanishes trivially. For $(k',l';k,l)=(+1,+1;-1,-1)$ one has $k'+l'-k-l=4$, $k'l'-kl=1-1=0$ (so $\Phi(t)$ drops out identically, since the Bell state connects two sectors of the same $\sigma_z^{(1)}\sigma_z^{(2)}$ parity), and $k+l-k'-l'=-4$, giving the remarkably simple result (Appendix~\ref{app:uncorrelated})
\begin{equation}
\rho_{14}^{\mathrm{unc}}(t)
=\tfrac12\,e^{-2i\omega_0t}\,e^{-4\Gamma(t)}.
\label{eq:rho14unc}
\end{equation}
Hence
\begin{equation}
\rho_S^{\mathrm{unc}}(t)=\frac12
\begin{pmatrix}
1&0&0&\rho_{14}^{\mathrm{unc}}(t)\\
0&0&0&0\\
0&0&0&0\\
\rho_{14}^{\mathrm{unc}}(t)^*&0&0&1
\end{pmatrix}.
\label{eq:rhounc_matrix}
\end{equation}
A single function $\Gamma(t)$ controls the entire coherence envelope; the dynamical phase $\Phi(t)$ never enters at all, for either the coherence or its conjugate, because the Bell state lives entirely inside a single $\sigma_z^{(1)}\sigma_z^{(2)}$ parity sector. This is a qualitative simplification relative to a generic product-state preparation, and it is the reason the Bell-state concurrence below takes such a compact form.

\subsection{Dynamics with initial correlations}
\label{sec:correlated}

We now let the interacting system and reservoir reach the joint equilibrium state before the qubit state is prepared. The initial system-reservoir state is
\begin{equation}
\rho(0) = \frac{\Omega\, e^{-\beta H}\,\Omega}{Z}, \quad \Omega=\ket\psi\bra\psi\otimes\mathds{1}_B,
\label{eq:corrinit}
\end{equation}
with $Z=\operatorname{Tr}_{S,B}\!\big[\Omega\, e^{-\beta H}\big]$ and $\ket\psi=\ket{\Phi^+}$ prepared by a local projective measurement at $t=0$. Because $H$, not $H_B$ alone, appears in the exponent, the reduced reservoir state is dressed by the qubit reservoir coupling before the projection, so Eq.~\eqref{eq:corrinit} genuinely differs from the factorized state of Eq.~\eqref{eq:factinit}. As shown in Appendix~\ref{app:correlated}, $H_S$ and $H_I$ are both functions of $J_z$, so $H$ is block diagonal in the $J_z$ eigenbasis. Sandwiching $e^{-\beta H}$ between $\bra\psi$ and $\ket\psi$ then collapses the system index completely, so Eq.~\eqref{eq:corrinit} turns out to be an exact product state $\rho(0)=\ket\psi\bra\psi\otimes(\mathcal O_B/Z)$: the qubits are exactly pure at $t=0$, and $\rho_S(0)$ is again given by Eq.~\eqref{eq:rho0}. What is not at equilibrium is the reduced reservoir state $\mathcal O_B/Z$, a $\ket\psi$ weighted mixture of displaced thermal states, one per $J_z$ sector. This is the entire content of the initial correlations in this model.

Evaluating the resulting Gaussian reservoir averages brings in a reorganization constant
\begin{equation}
\mathcal C \equiv 4\int_0^\infty \frac{J(\omega)}{\omega}\,d\omega,
\label{eq:Cdef}
\end{equation}
which is temperature independent and measures how far the reservoir relaxes around each displaced qubit sector, and an odd, real, temperature independent response function
\begin{equation}
\Lambda(t) \equiv 4\int_0^\infty \frac{J(\omega)}{\omega^2}\sin(\omega t)\,d\omega,
\label{eq:Lambdadef}
\end{equation}
which vanishes at $t=0$ and carries the entire new time dependence generated by the correlated preparation. For the Ohmic spectral density of Eq.~\eqref{eq:Jomega}, both are elementary: $\mathcal C=4g\omega_c$ and $\Lambda(t)=4g\arctan(\omega_ct)$.

Because $\ket{\Phi^+}$ only has support on the two sectors $(p,q)=(+1,+1)$ and $(-1,-1)$, each with equal weight $|\langle p,q|\psi\rangle|^2=\tfrac12$, the general correlation factor of Appendix~\ref{app:correlated} collapses to a two term sum, and the coupling constant $e^{\beta\mathcal C}$ cancels exactly between numerator and denominator. Writing $X(t)$ for the resulting complex multiplier of the $\rho_{14}$ coherence,
\begin{equation}
X(t) = \cos\bigl(4\Lambda(t)\bigr) - i\tanh(\beta\omega_0)\sin\bigl(4\Lambda(t)\bigr),
\label{eq:Xdef}
\end{equation}
so that
\begin{equation}
\rho_{14}^{\mathrm{corr}}(t) = \tfrac12\,e^{-2i\omega_0t}\,e^{-4\Gamma(t)}\,X(t).
\label{eq:rho14corr}
\end{equation}
The populations and the $\{\ket{01},\ket{10}\}$ block are completely unaffected by the correlations, exactly as for the factorized case, so
\begin{equation}
\rho_S^{\mathrm{corr}}(t)
=\frac12 
\begin{pmatrix}
1&0&0&\rho_{14}^{\mathrm{corr}}(t)\\
0&0&0&0\\
0&0&0&0\\
\rho_{14}^{\mathrm{corr}}(t)^*&0&0&1
\end{pmatrix}.
\label{eq:rhocorr_matrix}
\end{equation}
Since $|X(t)|\le1$ for all $t$ (Appendix~\ref{app:correlated}), with equality only at $t=0$ or when $\Lambda(t)$ is a multiple of $\pi/4$, the correlated coherence envelope $|X(t)|\,e^{-4\Gamma(t)}$ never exceeds the factorized envelope $e^{-4\Gamma(t)}$: initial correlations can only remove coherence, never add it. Setting $X(t)\to1$ recovers Eq.~\eqref{eq:rhounc_matrix} exactly, as any consistent generalization should.

\section{Entanglement dynamics with and without initial correlations}
\label{sec:entanglement}

The Wootters concurrence of a two qubit state $\rho$ is $C(\rho)=\max(0,\lambda_1-\lambda_2-\lambda_3-\lambda_4)$, where $\lambda_1\ge\cdots\ge\lambda_4\ge0$ are the square roots of the eigenvalues of $R=\rho\widetilde\rho$, and $\widetilde\rho=(\sigma_y\otimes\sigma_y)\rho^*(\sigma_y\otimes\sigma_y)$ is the spin flipped state~\cite{Wootters1998}. $C(\rho)=0$ for a separable state and $C(\rho)=1$ for a maximally entangled state. Both Eqs.~\eqref{eq:rhounc_matrix} and~\eqref{eq:rhocorr_matrix} have the ``$X$-state'' structure of a matrix $a$ on the diagonal and a single coherence $c(t)=\rho_{14}(t)$ off it, with equal populations $a=\rho_{11}=\rho_{44}=\tfrac12$. As shown in Appendix~\ref{app:uncorrelated}, equal populations make the state its own spin flip, $\widetilde\rho=\rho$, so $R=\rho^2$ has eigenvalues $(a+|c|)^2$, $(a-|c|)^2$, $0$, $0$, and Wootters' formula collapses to the remarkably compact result
\begin{equation}
C(t) = 2\,\bigl|\rho_{14}(t)\bigr|.
\label{eq:Cgeneral}
\end{equation}
Substituting Eqs.~\eqref{eq:rho14unc} and~\eqref{eq:rho14corr} (the unimodular phase $e^{\mp2i\omega_0t}$ drops out) gives the two central results of this paper:
\begin{align}
C_{\mathrm{unc}}(t) &= e^{-4\Gamma(t)},
\label{eq:Cunc}\\
C_{\mathrm{corr}}(t) &= e^{-4\Gamma(t)}\sqrt{1- \sech^2(\beta\omega_0)\sin^2\!\bigl(4\Lambda(t)\bigr)}.
\label{eq:Ccorr}
\end{align}
Equation~\eqref{eq:Cunc} is strictly positive for every finite $t$ since $\Gamma(t)$ stays finite at any finite time. A factorized Bell-state preparation never undergoes entanglement sudden death in this model; the concurrence decays smoothly and reaches zero only asymptotically, $t\to\infty$. Equation~\eqref{eq:Ccorr} is instead modulated by an oscillatory factor bounded between $|\tanh(\beta\omega_0)|$ and $1$. Because this factor reaches exact zero whenever $\cos(4\Lambda(t))=0$ together with $\tanh(\beta\omega_0)\to0$ (the high temperature regime $\beta\omega_0\ll1$), initial correlations are the only channel through which genuine entanglement sudden death and possible revival can occur for a Bell-state preparation coupled to a common Ohmic dephasing bath. At the opposite, zero temperature extreme, $\tanh(\beta\omega_0)\to1$ and Eq.~\eqref{eq:Ccorr} collapses back onto Eq.~\eqref{eq:Cunc}: the initial correlations become completely invisible in the concurrence as $T\to0$ (Appendix~\ref{app:correlated}). Correlations therefore act as a strictly detrimental, temperature-activated decay channel, on top of the always-present envelope $e^{-4\Gamma(t)}$, but they can never make the state more entangled than the factorized preparation at the same $t$.

\begin{figure}[t]
\centering
\includegraphics[width=0.98\linewidth]{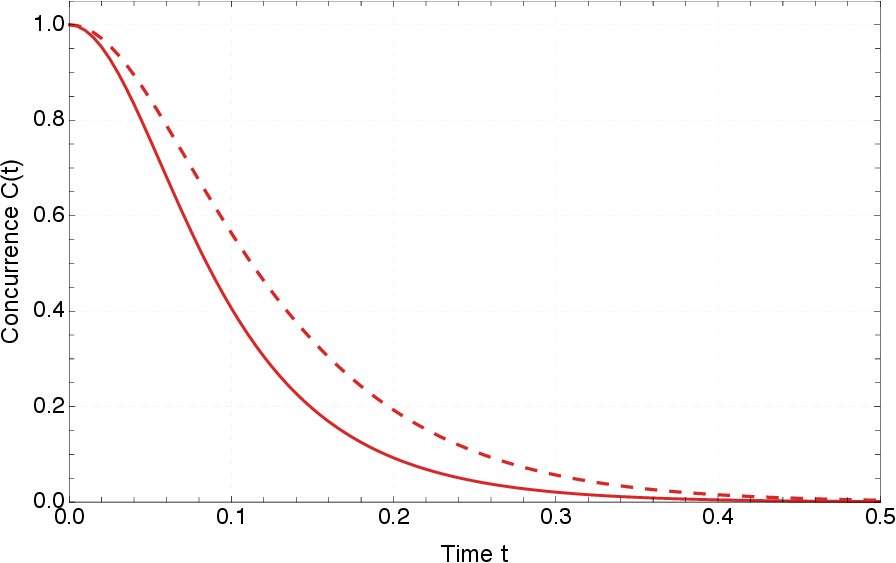}
\caption{(Color Online) Temporal evolution of the Concurrence $C(t)$, quantifying two-qubit entanglement dynamics in an exactly solvable two-qubit spin-boson model interacting with a common Ohmic bath for the factorized (dashed) and correlated (solid) Bell-state preparations. Chosen set of parameters are \rev{$\omega_0 = 1$, $\omega_c = 10$, $g=0.25$, $T=5$.}}
\label{fig:concurrence_corr}
\end{figure}

\rev{Figure~\ref{fig:concurrence_corr} shows the concurrence for the factorized and correlated Bell-state preparations at the comparatively strong coupling $g=0.25$ and high temperature $T=5$ (with $\omega_0=1$, $\omega_c=10$). Both curves decay quickly from $C(0)=1$ and have essentially vanished by $t\approx0.4$, and the correlated curve decays measurably faster than the factorized one, consistent with the bound $C_{\mathrm{corr}}(t)\le C_{\mathrm{unc}}(t)$ following from Eqs.~\eqref{eq:Cunc} and~\eqref{eq:Ccorr}. The growth of the gap between the $C_{\text{corr}}(t)$ and $C_{\text{unc}}(t)$ follows from Eqs.~\eqref{eq:Cunc} and~\eqref{eq:Ccorr}. The relative gap in which the decoherence factor $e^{-4\Gamma(t)}$ cancels, is}
\begin{equation}
{\Delta(t)\equiv\frac{C_{\mathrm{unc}}-C_{\mathrm{corr}}}{C_{\mathrm{unc}}}=1-\sqrt{1-\sech^2(\beta\omega_0)\,\sin^2\!\bigl(4\Lambda(t)\bigr)}.}
\label{eq:relgap}
\end{equation}
\rev{For $4\Lambda(t)=16g\arctan(\omega_ct)\ll1$ we may set $\sin x\simeq x$ and $\sqrt{1-y}\simeq1-y/2$, which gives}
\begin{align}
\Delta(t)&\simeq8\,\sech^2(\beta\omega_0)\,\Lambda(t)^2 \notag
\\
&=128\,g^2\,\sech^2(\beta\omega_0)\,\arctan^2(\omega_ct),
\label{eq:relgapquad}
\end{align}
\rev{so that}
\begin{equation}
{C_{\mathrm{unc}}-C_{\mathrm{corr}}\simeq128\,g^2\,e^{-4\Gamma(t)}\,\sech^2(\beta\omega_0)\,\arctan^2(\omega_ct).}
\label{eq:gapquad}
\end{equation}
\rev{The gap is quadratic hence superlinear in $g$. Two remarks delimit this statement. First, the expansion requires $16g\arctan(\omega_ct)\ll1$, i.e.\ $g\ll1/(8\pi)\approx0.04$ at long times. Second, $\Gamma\propto g$, so the absolute gap, Eq.~\eqref{eq:gapquad}, follows a pure $g^2$ law only while also $4\Gamma(t)\ll1$; at larger $g$ the factor $e^{-4\Gamma}$ suppresses it. The temperature dependence enters only through $\sech^2(\beta\omega_0)$, which vanishes as $T\to0$ (consistent with Appendix~\ref{app:correlated}) and tends to $1$ for $\beta\omega_0\ll1$. We checked Eq.~\eqref{eq:relgapquad} against the exact Eq.~\eqref{eq:Ccorr} at $t=1$, $\omega_0=1$, $\omega_c=10$, $T=1$: the log--log slope of $\Delta$ versus $g$ is $1.996$ for $g\in[10^{-3},10^{-2}]$ and $1.94$ for $g\in[10^{-3},5\times10^{-2}]$, and Eq.~\eqref{eq:relgapquad} reproduces the exact value to within $1.3\%$ at $g=0.01$ and $5\%$ at $g=0.02$.}

\begin{figure}[t]
\centering
\includegraphics[width=0.98\linewidth]{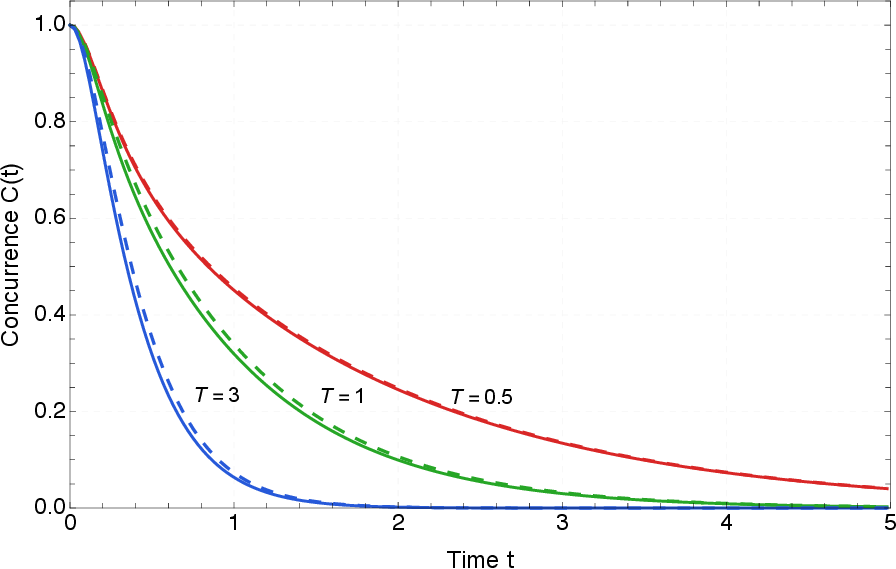}
\caption{(Color Online) Concurrence $C(t)$ for the factorized (dashed) and correlated (solid) Bell-state preparations at three temperatures, $T=0.5$ (red), $1.0$ (green), $3.0$(blue), with $g=0.1$, $\omega_0=1.0$, $\omega_c=10.0$. Higher temperature accelerates the loss of entanglement through $\coth(\beta\omega/2)$ in $\Gamma(t)$; at this coupling the correlated and factorized curves are barely distinguishable, consistent with the bound $C_{\mathrm{corr}}(t)\le C_{\mathrm{unc}}(t)$.}
\label{fig:concurrence_temp}
\end{figure}

Figure~\ref{fig:concurrence_temp} shows the concurrence at three representative temperatures, all six curves decay smoothly from $C(0)=1$ toward zero, with the decay setting in earlier and faster as the temperature is raised: the highest temperature curve has essentially vanished by $t\approx3$, while the lowest temperature curve retains sizeable concurrence out to $t=10$. This is exactly the behavior expected from Eq.~\eqref{eq:Gamma}: the thermal weight $\coth(\beta\omega/2)$ that drives $\Gamma(t)$ grows with temperature at every frequency, so $\Gamma(t)$, and hence the decay of $C_{\mathrm{unc}}(t)=e^{-4\Gamma(t)}$, grows faster at higher $T$. At this comparatively weak coupling ($g=0.05$) the correlated and factorized curves lie almost on top of one another at every temperature shown: the correction encoded in $\Lambda(t)$ stays small ($4\Lambda\lesssim0.3$ over the plotted window), so $\cos(4\Lambda)\approx1$ and $C_{\mathrm{corr}}(t)\approx C_{\mathrm{unc}}(t)$ throughout, consistent with the general bound $C_{\mathrm{corr}}(t)\le C_{\mathrm{unc}}(t)$ established above.

\begin{figure}[t]
\centering
\includegraphics[width=0.98\linewidth]{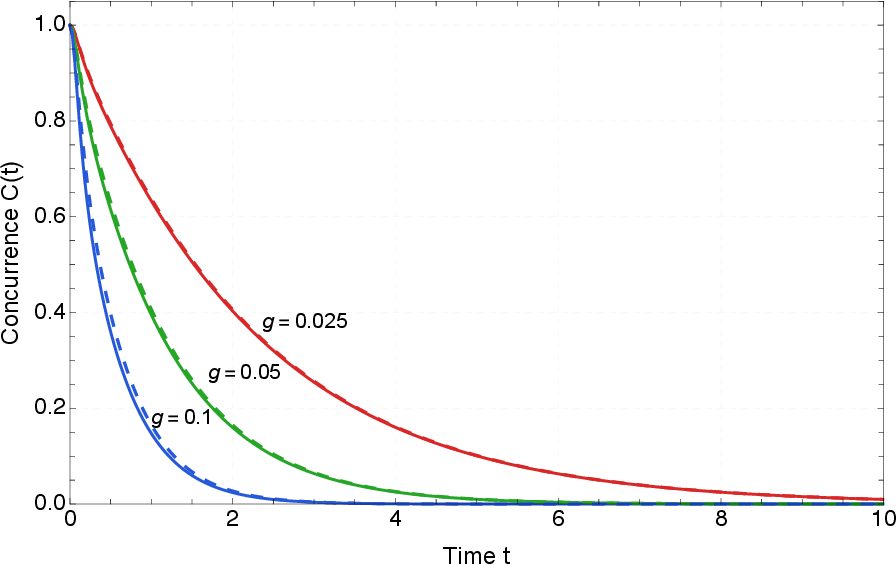}
\caption{(Color Online) Concurrence $C(t)$ for the factorized (dashed) and correlated (solid) Bell-state preparations at three qubit-reservoir coupling strengths, $g=0.1$ (red), $0.05$ (green), $0.025$(blue), with $T=1.5$, $\omega_0=1.0$, $\omega_c=10.0$. Stronger coupling accelerates the decay of $\Gamma(t)$ and hence of the entanglement envelope; the correlated curves again track the factorized ones very closely at these moderate couplings.}
\label{fig:concurrence_coupling}
\end{figure}

Figure~\ref{fig:concurrence_coupling} instead fixes $T=1.5$ and $\omega_c=10.0$ and sweeps the coupling strength, $g=0.1$ (red), $0.05$ (green), $0.025$(blue). Increasing $g$ increases the overall prefactor of $J(\omega)$ in Eq.~\eqref{eq:Gamma}, so $\Gamma(t)$, and with it the decay of the entanglement envelope $e^{-4\Gamma(t)}$, grows monotonically faster with $g$: the weakest coupling retains appreciable concurrence out to $t=10$, while the strongest coupling has essentially decayed to zero by $t\approx4$. As in Fig.~\ref{fig:concurrence_temp}, the correlated and factorized curves remain close together over this range of $g$, since the oscillatory correlation factor in Eq.~\eqref{eq:Ccorr} only departs from unity once $4\Lambda(t)=16g\arctan(\omega_ct)$ becomes an appreciable fraction of $\pi$; larger $g$ or $\omega_c$, or longer observation windows, would be needed to drive $\cos(4\Lambda)$ through zero and expose the genuine sudden-death and revival behavior implied by Eq.~\eqref{eq:Ccorr}.

\section{The role of control fields on the entanglement dynamics}
\label{sec:control}

Equations~\eqref{eq:Cunc}--\eqref{eq:Ccorr} show that the Bell-state concurrence decays irreversibly under pure dephasing, controlled entirely by the thermal function $\Gamma(t)$: for an Ohmic bath at finite temperature, $\Gamma(t)$ grows without bound as $t\to\infty$ because the $\coth(\beta\omega/2)$ weighting diverges at low frequency, so entanglement is eventually lost however weak the coupling. A natural question is whether the qubit pair can be actively steered to resist this decoherence. We adapt the strategy of dynamical decoupling: periodic, fast control pulses that average out the coupling between system and reservoir, in close analogy to Hahn echo and Carr-Purcell-Meiboom-Gill (CPMG) refocusing, as developed for qubit noise spectroscopy and demonstrated on solid-state and superconducting platforms~\cite{ViolaLloyd1998,ViolaKnillLloyd1999,NorrisPazSilvaViola2016,deLangeWangRisteDobrovitskiHanson2010,PokharelAnandFortmanLidar2018}. As emphasized in Ref.~\cite{AtherChaudhry2021}, control pulses and initial correlations act on physically distinct aspects of the probe-reservoir dynamics: the pulses reshape the reservoir spectral weight actually sampled by the probe, while the correlations reshape the reservoir state itself. It is therefore natural to ask whether the two ingredients combine constructively when the observable of interest is two-qubit entanglement.

We consider the symmetric application of $N$ instantaneous, equidistant $\pi$ pulses about the $x$ axis, applied simultaneously to both qubits, at times $t_j=j\tau$, $\tau=t/(N+1)$, $j=1,\dots,N$, throughout the interaction interval $[0,t]$~\cite{AtherChaudhry2021}. The collective pulse $P_\pi=\sigma_x^{(1)}\sigma_x^{(2)}$ satisfies $P_\pi^\dagger\sigma_z^{(i)}P_\pi=-\sigma_z^{(i)}$ ($i=1,2$), so every pulse inverts the sign of the qubit-reservoir coupling while leaving the bath operators untouched. Moving to the toggling frame is equivalent to the replacement $H_I(t')\to f_N(t')H_I(t')$, where the switching function
\begin{equation}
f_N(t') = 1+2\sum_{j=1}^N(-1)^j\Theta(t'-t_j) \equiv (-1)^{n(t')}
\label{eq:switch}
\end{equation}
alternates sign at every pulse, with $\Theta$ the Heaviside step function. Because $f_N(t')$ is a real $c$-number function, the exact truncation of the Magnus expansion established in Sec.~\ref{sec:model} is unaffected in form; only the integrals feeding $\alpha_r(t)$ and $\Phi(t)$ are modified (Appendix~\ref{app:uncorrelated}). Repeating the derivation of $\Gamma(t)$ and $\Lambda(t)$ with $f_N(t')$ inserted, and restricting to even $N$ so that the toggling frame coincides with the laboratory frame at the readout time $t$, gives the pulse-dressed dephasing function
\begin{equation}
\Gamma_N(t) = \int_0^\infty d\omega\,\frac{2J(\omega)}{\omega^2}\,F_N(\omega t)\,\coth\!\left(\frac{\beta\omega}{2}\right),
\label{eq:GammaN}
\end{equation}
with the filter function
\begin{equation}
F_N(\omega t) = 4\tan^2\!\left(\frac{\omega t}{2(N+1)}\right)\cos^2\!\left(\frac{\omega t}{2}\right),
\label{eq:FN}
\end{equation}
and the pulse-dressed correlation-response function
\begin{equation}
\Lambda_N(t) = 8g\int_0^\infty d\omega \frac{e^{-\omega/\omega_c}}{\omega} \tan\!\left(\frac{\omega t}{2(N+1)}\right)\cos^2\!\left(\frac{\omega t}{2}\right)
\label{eq:LambdaN}
\end{equation}
for the Ohmic spectral density. Both reduce identically to the unpulsed expressions, $F_0(\omega t)=4\sin^2(\omega t/2)$ and $\Lambda_{N=0}(t)=\Lambda(t)$, at $N=0$ (Appendix~\ref{app:uncorrelated}). At low frequency, $\omega\tau\ll1$, $F_N(\omega t)\propto1/(N+1)^2$: the filter function suppresses precisely the slow, secularly growing part of the bath spectrum responsible for the unbounded growth of $\Gamma(t)$, and does so more strongly as $N$ increases. The dynamical phase $\Phi_N(t)$ generalizes in the same way but, exactly as in Sec.~\ref{sec:model}, never enters the Bell-state coherence, for any $N$, so its closed form is not needed.

Because every step of the derivation in Sec.~\ref{sec:entanglement} depended on the density matrix only through the frozen, equal populations and a single coherence $c(t)$, both properties preserved verbatim under pulsing, the pulsed concurrence again collapses to $C^{(N)}(t)=2|\rho_{14}^{(N)}(t)|$, giving
\begin{align}
C_N^{\mathrm{unc}}(t) &= e^{-4\Gamma_N(t)},
\label{eq:CuncN}\\
C_N^{\mathrm{corr}}(t) &= e^{-4\Gamma_N(t)}\sqrt{1- \sech^2(\beta\omega_0)\sin^2\!\bigl(4\Lambda_N(t)\bigr)}.
\label{eq:CcorrN}
\end{align}
These have exactly the functional form of Eqs.~\eqref{eq:Cunc}--\eqref{eq:Ccorr}, with $\Gamma\to\Gamma_N$ and $\Lambda\to\Lambda_N$; since pulsing suppresses the low-frequency growth of $\Gamma_N(t)$ relative to the unpulsed $\Gamma(t)$, the entanglement is protected, decaying more slowly and surviving over a longer window than without control.

\begin{figure}[t]
\centering
\includegraphics[width=0.98\linewidth]{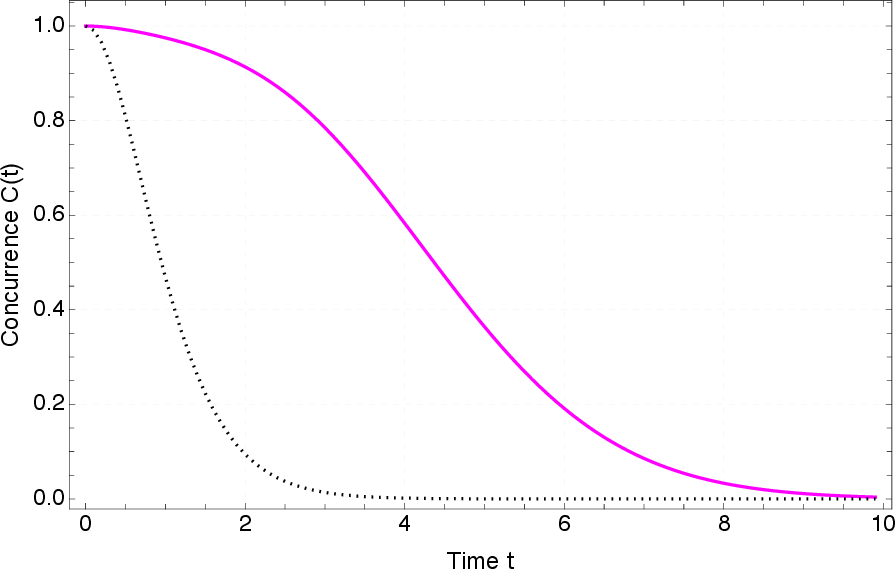}
\caption{(Color Online) Concurrence $C(t)$ without pulses ($N=0$, dotted) and under periodic $\pi$-pulse control (solid), for the correlated Bell-state preparations with $g=0.1$, $\omega_0=1$, $\omega_c=10$, $T=2.0$. Free evolution loses all entanglement by $t\approx2$--$4$, while the pulse-protected concurrence survives out to $t\approx9$--$10$: an extension of the entanglement lifetime by roughly a factor of four over this window.}
\label{fig:pulsed}
\end{figure}

Figure~\ref{fig:pulsed} compares the free evolution ($N=0$, dotted) against periodic $\pi$-pulse control (solid), at $g=0.1$, $\omega_0=1$, $\omega_c=1.0$, $T=2.0$. Without pulses, the entanglement decays essentially completely by $t\approx2$--$4$. With pulse control the decay instead sets in only after an extended plateau and the concurrence survives out to $t\approx9$--$10$, roughly quadrupling the window over which the qubits stay entangled. The correlated and factorized curves remain visually indistinguishable throughout, both with and without pulsing, at this coupling strength and temperature: the correction terms in Eqs.~\eqref{eq:CuncN}--\eqref{eq:CcorrN} stay small relative to the dominant envelope $e^{-4\Gamma_N(t)}$ over the plotted range.

\begin{figure}[t]
\centering
\includegraphics[width=0.98\linewidth]{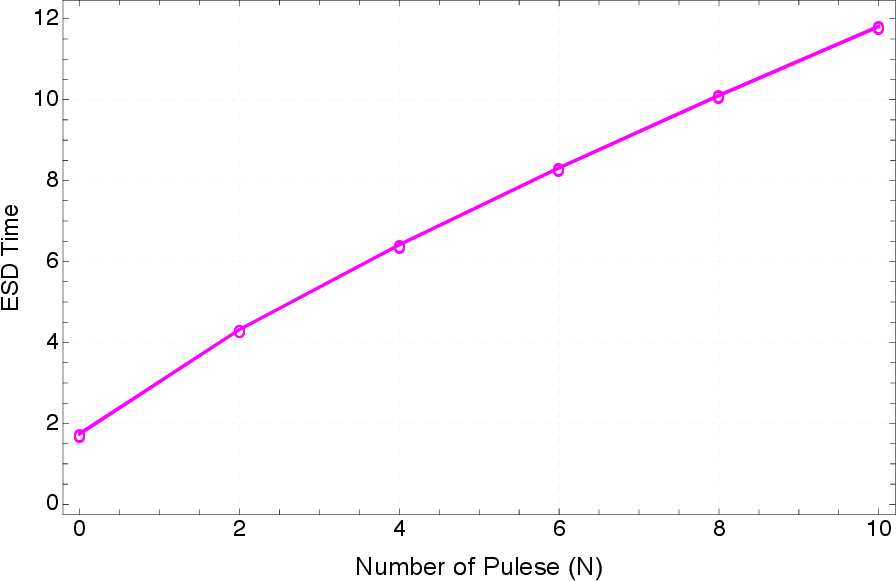}
\caption{(Color Online) Entanglement death time $t_{\rm death}$, defined as the first time the concurrence falls below the numerical threshold $C(t)<10^{-3}$, as a function of the (even) number of pulses $N\in\{0,2,4,6,8,10\}$, for the factorized and correlated preparations (overlapping), at $g=0.15$, $\omega_0=\omega_c=1.0$, $T=5.0$. $t_{\rm death}$ grows monotonically with $N$, from $t_{\rm death}\approx1.7$ at $N=0$ to $t_{\rm death}\approx11.9$ at $N=10$, delivering close to a seven-fold extension of the entanglement lifetime over the range of pulse numbers studied.}
\label{fig:esd_vs_N}
\end{figure}

Figure~\ref{fig:esd_vs_N} quantifies this protection more systematically, using a numerically defined death time $t_{\rm death}$ (the first instant at which $C(t)$ falls below the threshold $10^{-3}$, since the factorized concurrence, Eq.~\eqref{eq:CuncN}, is strictly positive and only reaches zero asymptotically) as a function of the even pulse number $N\in\{0,2,4,6,8,10\}$, at $g=0.15$, $\omega_0=\omega_c=1.0$, $T=5.0$. The death time grows monotonically and close to linearly with $N$ over this range, from $t_{\rm death}\approx1.7$ at $N=0$ to $t_{\rm death}\approx11.9$ at $N=10$: nearly a seven-fold extension of the entanglement lifetime. The correlated and factorized curves track one another essentially exactly at every $N$ probed, confirming that the pulse-sequence protection mechanism, encoded entirely in the suppression of $\Gamma_N(t)$ by the filter function $F_N(\omega t)$, acts essentially independently of, and on top of, the correlation-induced physics of Sec.~\ref{sec:entanglement}.

\section{Conclusion}
\label{sec:conclusion}

We have solved exactly for the entanglement dynamics of two qubits, prepared in the maximally entangled Bell state $\ket{\Phi^+}$, coupled identically to a common Ohmic dephasing reservoir, for both a factorized initial state and a correlated, projected thermal initial state in which the qubits and reservoir first reach a joint equilibrium and the qubit state is prepared only afterward by a local measurement. In both cases the pure dephasing structure of the interaction guarantees exact solvability to all orders in the coupling, and the equal-population, single-coherence structure of the Bell-state preparation collapses the Wootters concurrence to the remarkably simple form $C(t)=2|\rho_{14}(t)|$ for both preparations, with and without pulse control.

For the factorized state we found $C_{\mathrm{unc}}(t)=e^{-4\Gamma(t)}$: strictly positive at every finite time, so a factorized Bell-state preparation never undergoes true entanglement sudden death in this model, only asymptotic decay. For the correlated state, an additional, purely correlation-induced oscillatory factor modulates this envelope, bounded between $|\tanh(\beta\omega_0)|$ and $1$; it is this factor, and only this factor, that can drive the concurrence to exact zero at a finite time, in the high-temperature regime $\beta\omega_0\ll1$. Initial correlations therefore play a qualitatively distinct, and strictly detrimental, role relative to the factorized case: at the couplings and temperatures probed numerically here the two preparations remain close, but the correlated preparation is the only one of the two capable of genuine sudden death. We further showed that a symmetric sequence of periodic $\pi$ pulses, a two-qubit dynamical decoupling protocol, extends the entanglement lifetime substantially, by close to a factor of seven over the range of even pulse numbers studied, acting as an essentially independent knob on top of, rather than in place of, the correlation-induced physics.

These results complement the spin environment analysis of Ref.~\cite{MajeedChaudhry2019}, where the sign and magnitude of the initial correlation effect depend sensitively on microscopic environment details. Here, for a common bosonic dephasing environment and a maximally entangled initial preparation, the effect of initial correlations is structurally simple: it can only suppress the concurrence relative to the factorized envelope, and it is the sole mechanism by which the entanglement of this particular preparation can be exactly and permanently extinguished at a finite time. Our dynamical decoupling results also extend the single-probe program of Ref.~\cite{AtherChaudhry2021} to a genuinely two-body observable.

\begin{acknowledgments}
AR Mirza thanks the Higher Education Commission for support through IPFP (Interim Placement for fresh Phds) fellowship.
\end{acknowledgments}

\appendix

\section{Exact Bell-state dynamics without initial correlations}
\label{app:uncorrelated}

This appendix gives the full derivation of the results quoted in Secs.~\ref{sec:uncorrelated}, \ref{sec:entanglement}, and \ref{sec:control} for the factorized initial state, Eq.~\eqref{eq:factinit}.

\subsection{Interaction picture and the Magnus expansion}

Moving to the interaction picture with respect to $H_0=H_S+H_B$ via $U_0(t)=e^{i(H_B+H_S)t}$, only the bath operators pick up dynamical phases, $b_r(t)=b_re^{-i\omega_rt}$, so that
\begin{equation}
H_I(t) = J_z\sum_r\left(g_r^*b_re^{-i\omega_rt}+g_rb_r^\dagger e^{i\omega_rt}\right).
\label{eq:HIt}
\end{equation}
The propagator $U_I(t)$ satisfies $i\dot U_I(t)=H_I(t)U_I(t)$, and the Magnus expansion reorganizes the associated Dyson series into a single exponential, $U_I(t)=\exp[\sum_iA_i(t)]$, with
\begin{align}
A_1(t) &= -i\int_0^tH_I(t_1)\,dt_1,\\
A_2(t) &= -\frac12\int_0^tdt_1\int_0^{t_1}dt_2\,\big[H_I(t_1),H_I(t_2)\big].
\end{align}
Because $H_I(t)$ is linear in the bath ladder operators, $[H_I(t_1),H_I(t_2)]$ is a $c$-number (proportional to the identity on the bath and diagonal in $\sigma_z^{(1)}\sigma_z^{(2)}$), so every Magnus term beyond second order vanishes identically, $A_3=A_4=\cdots=0$: the expansion is an exact resummation of the Dyson series, valid at any coupling strength. Substituting Eq.~\eqref{eq:HIt} into $A_1$ and integrating gives
\begin{equation}
A_1(t) = \frac12J_z\sum_r\left[\alpha_r(t)b_r^\dagger-\alpha_r^*(t)b_r\right],
\end{equation}
with 
\begin{equation}
\alpha_r(t)=\frac{2g_r(1-e^{i\omega_rt})}{\omega_r},
\end{equation}
i.e.\ $\tfrac12J_z$ times a bath displacement generator. For the commutator, using $[b_r,b_{r'}^\dagger]=\delta_{rr'}$ and $J_z^2=2(\mathds1+\sigma_z^{(1)}\sigma_z^{(2)})$,
\begin{align}
\big[H_I(t_1),H_I(t_2)\big] 
&= 4i\left(\mathds1+\sigma_z^{(1)}\sigma_z^{(2)}\right)\notag \\
&\times \sum_r|g_r|^2\sin\!\big[\omega_r(t_2-t_1)\big],
\end{align}
a $c$-number times a qubit-diagonal operator built from $\sigma_z^{(1)}\sigma_z^{(2)}$, not from $J_z$ alone. Carrying out the double time integral gives
\begin{equation}
A_2(t) = -\frac{i}{2}\left(\mathds1+\sigma_z^{(1)}\sigma_z^{(2)}\right)\Phi(t),
\end{equation}
with
\begin{equation}
\Phi(t) = \sum_r\frac{4|g_r|^2}{\omega_r^2}\Big[\sin(\omega_rt)-\omega_rt\Big].
\end{equation}
$\Phi(t)$ is real and multiplies the eigenvalue of $\mathds1+\sigma_z^{(1)}\sigma_z^{(2)}$, which for a coherence between $(k,l)$ and $(k',l')$ contributes a net phase proportional to $kl-k'l'$.

\subsection{Exact propagator and reduced density matrix}

Combining $U_0(t)=e^{-i\frac{\omega_0}{2}J_zt}e^{iH_Bt}$ with $U_I(t)=\exp[A_1(t)+A_2(t)]$, and writing $P_{kl,k'l'}\equiv\ket{k,l}\bra{k',l'}$, the Heisenberg-picture projector is
\begin{align}
P_{kl,k'l'}(t) 
&= e^{i\frac{\omega_0}{2}(k+l-k'-l')t}\notag \\
&\times e^{i\frac{\Phi(t)}{2}(kl-k'l')}\;e^{\widetilde R_{kl,k'l'}(t)}\;P_{kl,k'l'},
\end{align}
with $\widetilde R_{kl,k'l'}(t)=\sum_r[\widetilde\alpha_r(t)b_r^\dagger-\widetilde\alpha_r^*(t)b_r]$ and $\widetilde\alpha_r(t)=\tfrac12(k+l-k'-l')\alpha_r(t)$. Tracing over the bath with $\rho_B=e^{-\beta H_B}/Z_B$, the bath average is Gaussian, $\langle e^X\rangle=e^{\frac12\langle X^2\rangle}$ for $X$ linear in bosonic operators with $\langle X\rangle=0$, giving
\begin{equation}
\big\langle e^{\widetilde R_{kl,k'l'}(t)}\big\rangle_{\rho_B} = \exp\!\left[-\frac14(k+l-k'-l')^2\Gamma(t)\right],
\end{equation}
using $2n_r+1=\coth(\beta\omega_r/2)$, with $\Gamma(t)$ as in Eq.~\eqref{eq:Gamma}. Combining these pieces reproduces Eq.~\eqref{eq:rhoSfinal}, and evaluating it for $\rho_S(0)=\ket{\Phi^+}\bra{\Phi^+}$ [Eq.~\eqref{eq:rho0}] at the single nonzero coherence $(k',l';k,l)=(+1,+1;-1,-1)$: since $k=l=-1\Rightarrow kl=1$ and $k'=l'=1\Rightarrow k'l'=1$, the phase coefficient $k'l'-kl=0$ vanishes identically, giving Eq.~\eqref{eq:rho14unc}.

\subsection{Diagonalization and the closed-form concurrence}

The reduced density matrix, Eq.~\eqref{eq:rhounc_matrix}, is block diagonal: a $2\times2$ block on $\{\ket{00},\ket{11}\}$ with diagonal entries $a=\tfrac12$ and off-diagonal entry $c(t)=\rho_{14}(t)$, and a zero block on $\{\ket{01},\ket{10}\}$. The $2\times2$ block diagonalizes as
\begin{align}
\ket{v_\pm} 
&= \tfrac{1}{\sqrt2}\bigl(\ket{00}\pm e^{-i\theta(t)}\ket{11}\bigr),
\\
\mu_\pm 
&= a\pm|c(t)|,
\\
c(t)
&=|c(t)|e^{i\theta(t)},
\label{eq:eigvecs}
\end{align}
together with $\ket{01},\ket{10}$ at eigenvalue $0$. Positivity of $\rho_S(t)$ requires $|c(t)|\le a=\tfrac12$, which holds since $|c(t)|=\tfrac12e^{-4\Gamma(t)}\le\tfrac12$ for all $t$. For the spin flip, with $Y\equiv\sigma_y\otimes\sigma_y$ the antidiagonal matrix $Y\ket{00}=-\ket{11}$, $Y\ket{01}=\ket{10}$, $Y\ket{10}=\ket{01}$, $Y\ket{11}=-\ket{00}$, a direct computation for a general block-diagonal state $\rho=\mathrm{diag\ block}(a,c;c^*,d)\oplus0$ gives
\begin{equation}
\widetilde\rho = Y\rho^*Y =
\begin{pmatrix}
d&0&0&c\\0&0&0&0\\0&0&0&0\\c^*&0&0&a
\end{pmatrix}.
\label{eq:rhotilde_general}
\end{equation}
Because $a=d=\tfrac12$ for the Bell-state preparation (both with and without correlations, Sec.~\ref{sec:model}), $\widetilde\rho=\rho$ exactly: the state is its own spin flip. Hence $R=\rho\widetilde\rho=\rho^2$, whose nontrivial $2\times2$ block has eigenvalues $(a\pm|c|)^2$, giving ordered square roots $\lambda_1=a+|c|$, $\lambda_2=a-|c|$, $\lambda_3=\lambda_4=0$ and
\begin{equation}
C(t) = \max\bigl(0,\ (a+|c|)-(a-|c|)\bigr) = 2|c(t)|,
\end{equation}
which is Eq.~\eqref{eq:Cgeneral}; the $\max(0,\cdot)$ is automatically saturated at the positive branch since $|c(t)|\ge0$. Substituting $c(t)=\rho_{14}^{\mathrm{unc}}(t)=\tfrac12e^{-2i\omega_0t}e^{-4\Gamma(t)}$ gives $C_{\mathrm{unc}}(t)=e^{-4\Gamma(t)}$, Eq.~\eqref{eq:Cunc}; since $\Gamma(t)\ge0$ is finite for every finite $t$, $C_{\mathrm{unc}}(t)>0$ strictly, and no entanglement sudden death occurs from the factorized preparation.

\subsection{Pulsed generalization}

Under the periodic $\pi$-pulse sequence of Sec.~\ref{sec:control}, the substitution $H_I(t')\to f_N(t')H_I(t')$ leaves the structure of the Magnus expansion unchanged, since $f_N(t')$ is a real $c$-number function and $A_3=A_4=\cdots=0$ continues to hold exactly. Only the coupling integral feeding $A_1$ is modified, $\alpha_r(t)\to\alpha_r^{(N)}(t)=-2ig_ry_N(\omega_r,t)$ with $y_N(\omega,t)=\int_0^tf_N(t')e^{i\omega t'}dt'$. For the symmetric, equally spaced sequence $t_j=j\tau$, $\tau=t/(N+1)$, splitting the integral into $N+1$ alternating segments and summing the resulting finite geometric series gives, for even $N$,
\begin{align}
y_N(\omega,t) 
&= \frac{\tan(\omega\tau/2)}{\omega}\bigl(1+e^{i\omega t}\bigr),
\\
|y_N(\omega,t)|^2 
&= \frac{4\tan^2(\omega\tau/2)}{\omega^2}\cos^2\!\Bigl(\frac{\omega t}2\Bigr),
\\
\mathrm{Re}\,y_N(\omega,t) 
&= \frac{2\tan(\omega\tau/2)}{\omega}\cos^2\!\Bigl(\frac{\omega t}2\Bigr).
\end{align}
At $N=0$ ($\tau=t$) these reduce to $|y_0|^2=4\sin^2(\omega t/2)/\omega^2$ and $\mathrm{Re}\,y_0=\sin(\omega t)/\omega$, reproducing the unpulsed integrals used to build $\Gamma(t)$ and $\Lambda(t)$ exactly. Substituting $\omega^2|y_N|^2\equiv F_N(\omega t)$ into $\Gamma_N(t)=\tfrac12\sum_r|\alpha_r^{(N)}(t)|^2(2n_r+1)$ gives Eq.~\eqref{eq:GammaN}--\eqref{eq:FN}, and substituting $\mathrm{Re}\,y_N$ into the pulsed generalization of the correlation-response integral gives Eq.~\eqref{eq:LambdaN}. Because the dynamical phase $\Phi_N(t)$ again multiplies only $kl-k'l'$, which vanishes identically for the Bell-state coherence for every pulse sequence, it never enters the pulsed dynamics, and the entire derivation of Sec.~\ref{sec:entanglement} carries over verbatim with $\Gamma\to\Gamma_N$, $\Lambda\to\Lambda_N$, giving Eqs.~\eqref{eq:CuncN}--\eqref{eq:CcorrN}. As a benchmark, setting $N=0$ in Eqs.~\eqref{eq:CuncN}--\eqref{eq:CcorrN} reproduces Eqs.~\eqref{eq:Cunc}--\eqref{eq:Ccorr} exactly, to the resolution of the numerical time grid used in Figs.~\ref{fig:pulsed}--\ref{fig:esd_vs_N}.

\section{Exact Bell-state dynamics with initial correlations}
\label{app:correlated}

This appendix gives the full derivation of the results quoted in Sec.~\ref{sec:correlated} for the correlated, projected thermal initial state, Eq.~\eqref{eq:corrinit}.

\subsection{The projected thermal state is an exact product state}

Because $H_S$ and $H_I$ are both built out of $J_z=\sigma_z^{(1)}+\sigma_z^{(2)}$, the full Hamiltonian $H$ is block diagonal in the $J_z$ eigenbasis $\ket{p,q}$ ($p,q=\pm1$). Within the sector of fixed $p+q$ it acts purely on the bath,
\begin{align}
H\big|_{p+q} 
&= \frac{\omega_0}{2}(p+q) + H_B^{(p+q)},
\\
H_B^{(\lambda)} 
&\equiv H_B+\lambda\sum_r\bigl(g_r^*b_r+g_rb_r^\dagger\bigr).
\end{align}
Sandwiching $e^{-\beta H}$ between $\bra\psi$ and $\ket\psi$, with $\ket\psi=\ket{\Phi^+}$ supported only on $(p,q)=(+1,+1)$ and $(-1,-1)$, each with weight $\tfrac12$, collapses the system index completely, leaving a pure bath operator
\begin{equation}
\bra\psi e^{-\beta H}\ket\psi = \tfrac12e^{-\beta\omega_0}e^{-\beta H_B^{(1)}} + \tfrac12e^{\beta\omega_0}e^{-\beta H_B^{(-1)}} \equiv \mathcal O_B,
\end{equation}
so that $\Omega e^{-\beta H}\Omega=\ket\psi\bra\psi\otimes\mathcal O_B$, i.e.\ Eq.~\eqref{eq:corrinit} is an exact product state, $\rho(0)=\ket\psi\bra\psi\otimes(\mathcal O_B/Z)$: the qubits are exactly pure, $\rho_S(0)=\ket{\Phi^+}\bra{\Phi^+}$, at $t=0$. What is not equilibrium is the bath factor $\mathcal O_B/Z$, a $\ket\psi$-weighted mixture of displaced thermal states, one per $J_z$ sector.

\subsection{Bath average under the dressed state}

Completing the square in $H_B^{(\lambda)}$ (a displaced-oscillator shift, $B_r\equiv b_r+\lambda g_r/\omega_r$) gives the exact identities
\begin{align}
\operatorname{Tr}_B\bigl[e^{-\beta H_B^{(\lambda)}}\bigr] 
&= Z_Be^{\beta\lambda^2\mathcal C/8},
\\
\operatorname{Tr}_B\bigl[e^{-\beta H_B^{(\lambda)}}e^{\widetilde R_{kl,k'l'}(t)}\bigr] 
&= Z_Be^{\beta\lambda^2\mathcal C/8}\,e^{-\frac14(k+l-k'-l')^2\Gamma(t)} \notag
\\
& \times e^{i\lambda(k+l-k'-l')\Lambda(t)/4},
\end{align}
with $\mathcal C$, $\Lambda(t)$ as in Eqs.~\eqref{eq:Cdef}--\eqref{eq:Lambdadef}. Combining this with the weighted sum over the two $J_z$ sectors that overlap $\ket{\Phi^+}$ gives, for the $(+1,+1;-1,-1)$ matrix element (for which $k+l-k'-l'=-4$),
\begin{align}
&\text{Denominator} = \tfrac12e^{-\beta\omega_0}e^{\beta\mathcal C} + \tfrac12e^{\beta\omega_0}e^{\beta\mathcal C} = e^{\beta\mathcal C}\cosh(\beta\omega_0),\\
&\text{Numerator} 
= \tfrac12e^{-\beta\omega_0}e^{\beta\mathcal C}e^{4i\Lambda(t)} + \tfrac12e^{\beta\omega_0}e^{\beta\mathcal C}e^{-4i\Lambda(t)}\notag \\
&= e^{\beta\mathcal C}\Bigl[\cosh(\beta\omega_0)\cos(4\Lambda(t)) - i\sinh(\beta\omega_0)\sin(4\Lambda(t))\Bigr].
\end{align}
The coupling constant $e^{\beta\mathcal C}$ cancels between numerator and denominator, leaving the correlation multiplier
\begin{equation}
X(t) = \cos\bigl(4\Lambda(t)\bigr) - i\tanh(\beta\omega_0)\sin\bigl(4\Lambda(t)\bigr),
\end{equation}
which is Eq.~\eqref{eq:Xdef}, giving $\rho_{14}^{\mathrm{corr}}(t)=\rho_{14}^{\mathrm{unc}}(t)\,X(t)$, Eq.~\eqref{eq:rho14corr}. Repeating the calculation for the conjugate element $(-1,-1;+1,+1)$ gives $X(t)^*$, consistent with Hermiticity. Note $X(0)=1$ (since $\Lambda(0)=0$), so $\rho_S^{\mathrm{corr}}(0)=\rho_S(0)$ of Eq.~\eqref{eq:rho0} is recovered exactly, and $X(t)\to1$ whenever $\tanh(\beta\omega_0)\to1$ and $\Lambda(t)\to0$, or more generally as $\beta\to\infty$: since $\tanh(\beta\omega_0)\to1$ for any $\omega_0\ne0$ at $T\to0$, Eq.~\eqref{eq:Xdef} gives $|X(t)|\to\sqrt{\cos^2(4\Lambda)+\sin^2(4\Lambda)}=1$, so $C_{\mathrm{corr}}(t)\to C_{\mathrm{unc}}(t)$: initial correlations become invisible in the concurrence at zero temperature.

\subsection{Bound on $|X(t)|$ and the origin of correlated sudden death}

Since $|X(t)|^2=\cos^2(4\Lambda)+\tanh^2(\beta\omega_0)\sin^2(4\Lambda) = 1-\bigl[1-\tanh^2(\beta\omega_0)\bigr]\sin^2(4\Lambda)\le1$ for all $t$, with equality only at $\Lambda(t)=n\pi/4$, the correlated concurrence obeys $C_{\mathrm{corr}}(t)\le C_{\mathrm{unc}}(t)=e^{-4\Gamma(t)}$ at every instant: correlations can only equal or reduce the concurrence relative to the factorized envelope, never exceed it. $C_{\mathrm{corr}}(t)$ reaches exactly zero only when $\cos(4\Lambda(t))=0$ and $\tanh(\beta\omega_0)\to0$ simultaneously, i.e.\ in the high-temperature limit $\beta\omega_0\ll1$; this is the unique mechanism, absent from the factorized preparation, by which the Bell-state entanglement can undergo true sudden death at a finite time.

\subsection{Pulsed generalization}

Under the periodic $\pi$-pulse sequence of Sec.~\ref{sec:control}, the reorganization constant $\mathcal C=4g\omega_c$ is a static, $t=0$ equilibrium quantity fixed by the preparation, and is completely unaffected by pulses applied for $t>0$. The response function $\Lambda(t)$, by contrast, acquires the pulse-dressed form $\Lambda_N(t)$ of Eq.~\eqref{eq:LambdaN}, obtained by the same substitution $y_0\to y_N$ used for $\Gamma_N(t)$ in Appendix~\ref{app:uncorrelated}. The correlated concurrence under pulsed control, Eq.~\eqref{eq:CcorrN}, then follows from the same two-term $(p,q)=(\pm1,\pm1)$ collapse as above, with $\Gamma\to\Gamma_N$ and $\Lambda\to\Lambda_N$ throughout; setting $N=0$ reproduces the unpulsed correlated concurrence of Sec.~\ref{sec:correlated} exactly.

\twocolumngrid

\bibliographystyle{apsrev4-2}
\bibliography{references}

\end{document}